\RequirePackage[l2tabu, orthodox]{nag}
\documentclass[11pt]{article}

\usepackage[T1]{fontenc}
\usepackage{tcolorbox}

\usepackage{soul} 

\usepackage{garamond}  
\renewcommand{\rmdefault}{ugm}

\usepackage[bitstream-charter]{mathdesign}
\usepackage{amsmath}
\usepackage[scaled=0.92]{PTSans}

\usepackage[
  paper  = letterpaper,
  left   = 1.65in,
  right  = 1.65in,
  top    = 1.0in,
  bottom = 1.0in,
  ]{geometry}

\usepackage[dvipsnames]{xcolor}
\definecolor{shadecolor}{gray}{0.9}

\usepackage[final,expansion=alltext]{microtype}
\usepackage[english]{babel}
\usepackage[parfill]{parskip}
\usepackage{afterpage}
\usepackage{framed}

{\endMakeFramed}

\DeclareRobustCommand{\parhead}[1]{\textbf{#1}~}

\usepackage{lineno}

\usepackage{ragged2e}

\newcounter{parcount}

\usepackage{graphicx}
\usepackage[labelfont=bf]{caption}

\usepackage{booktabs}
\usepackage{multirow}

\usepackage[algoruled]{algorithm2e}
\usepackage{listings}
\usepackage{fancyvrb}
\fvset{fontsize=\normalsize}

\usepackage{natbib}

\usepackage[acronym,nowarn]{glossaries}
\makeglossaries

\definecolor{tangerine}{rgb}{0.95, 0.52, 0.0}
\definecolor{palebrown}{rgb}{0.6, 0.46, 0.33}
\definecolor{peru}{rgb}{0.8, 0.52, 0.25}
\usepackage[colorlinks,linktoc=all]{hyperref}
\usepackage[all]{hypcap}
\hypersetup{citecolor=peru}
\hypersetup{linkcolor=peru}
\hypersetup{urlcolor=peru}

\usepackage[nameinlink]{cleveref}
\creflabelformat{equation}{#1#2#3}
\crefname{equation}{eq.}{eqs.}  
\Crefname{equation}{Eq.}{Eqs.}

\lstdefinestyle{mystyle}{
    commentstyle=\color{OliveGreen},
    keywordstyle=\color{BurntOrange},
    numberstyle=\tiny\color{black!60},
    stringstyle=\color{MidnightBlue},
    basicstyle=\ttfamily,
    breakatwhitespace=false,
    breaklines=true,
    captionpos=b,
    keepspaces=true,
    numbers=left,
    numbersep=5pt,
    showspaces=false,
    showstringspaces=false,
    showtabs=false,
    tabsize=2
}
\usepackage[colorinlistoftodos,
           prependcaption,
           textsize=small,
           backgroundcolor=yellow,
           linecolor=lightgray,
           bordercolor=lightgray]{todonotes}

\DeclareRobustCommand{\parhead}[1]{\textbf{#1}~}

\usepackage{amsthm}  

\usepackage{bm}

\usepackage[colorinlistoftodos,
           prependcaption,
           textsize=small,
           backgroundcolor=yellow,
           linecolor=lightgray,
           bordercolor=lightgray]{todonotes}

\usepackage{graphicx}
\usepackage{caption}
\usepackage{subcaption}
\usepackage{wrapfig}

\usepackage{booktabs}
\usepackage{arydshln} 
\usepackage{multirow}
\usepackage{nicematrix}

\usepackage{listings}
\usepackage{fancyvrb}
\fvset{fontsize=\normalsize}

\usepackage[nameinlink]{cleveref}
\creflabelformat{equation}{#1#2#3}
\crefname{equation}{eq.}{eqs.}  
\Crefname{equation}{Eq.}{Eqs.}

\usepackage{listings}
\lstdefinestyle{alp_style}{
    commentstyle=\color{OliveGreen},
    numberstyle=\tiny\color{black!60},
    stringstyle=\color{BrickRed},
    basicstyle=\ttfamily\scriptsize,
    breakatwhitespace=false,
    breaklines=true,
    captionpos=b,
    keepspaces=true,
    numbers=none,
    numbersep=5pt,
    showspaces=false,
    showstringspaces=false,
    showtabs=false,
    tabsize=2
}

\usepackage{capt-of}

\usepackage{lipsum}

\newtheorem{theorem}{Theorem}[section]

\theoremstyle{remark}
\newtheorem*{lemma*}{Lemma}

\newcommand{\bp}{\bm{p}}

\newcommand{\bK}{\bm{K}}

\newcommand{\by}{\bm{y}}

\newcommand{\bx}{\bm{x}}
\newcommand{\bs}{\bm{s}}

\newcommand{\bt}{\bm{t}}

\usepackage{authblk}
\usepackage{enumitem} 
\usepackage{algorithmic}
\usepackage{tabularx}
\usepackage{booktabs}
\usepackage{pifont}
\usepackage[T1]{fontenc}
\usepackage{multirow}
\usepackage{threeparttable}

\title{\textbf{A Unified and Predictive Measure of\\ Functional Diversity}}

\author[1, 3]{Adji Bousso Dieng}
\author[2, 3]{Amey P. Pasarkar}
\affil[1]{Department of Computer Science, Princeton University}
\affil[2]{Lewis-Sigler Institute For Integrative Genomics, Princeton University}
\affil[3]{\href{https://vertaix.princeton.edu/}{Vertaix}}

\begin{document}
\maketitle

\begin{abstract}
\noindent Despite the critical role of functional diversity (FD) in understanding ecological systems and processes, its robust quantification remains a significant challenge. A long-held view in the field is that it is not possible to capture its three facets---functional richness, functional divergence, and functional evenness---in a single index. This perspective has prompted recent proposals for FD measurement to use three separate indices, one for each aspect. Here, we challenge this paradigm by demonstrating that the probability-weighted Vendi Score (pVS), first introduced by \citet{friedman2023vendi}, can serve as a powerful functional diversity index that can capture its three facets. We adapt pVS to functional ecology by defining it as the exponential of the Rényi entropy of the eigenvalues of the abundance-weighted trait similarity matrix. This formulation allows pVS to be applicable at any biological level. It can be defined at the species level, at which most existing FD metrics are defined, and at the individual level to naturally incorporate intraspecific trait variation (ITV) when detailed data are available. We theoretically and empirically demonstrate the robustness of pVS. We first mathematically prove it satisfies several essential desiderata for FD metrics, including invariance to functional redundancy, set monotonicity, distance monotonicity, and concavity. We then show that pVS consistently exhibits the expected ground-truth behavior on simulated ecosystem scenarios under which many FD metrics fail. By integrating abundances and trait similarities within a single, theoretically sound framework, pVS provides a generally applicable index for ecology.\\ 

\end{abstract}

\section{Introduction}
The robust and comprehensive quantification of functional diversity (FD) has remained a persistent challenge in ecology. FD is commonly conceptualized through three distinct facets: functional richness, functional divergence, and functional evenness~\citep{villeger2008, Schmera2023}. Functional richness is the number of distinct functional types or the volume of functional trait space occupied by the community. It reflects the variety of functional roles present. Functional divergence on the other hand describes how functionally dissimilar individuals or species are from each other and how evenly they are spread across the functional trait space. Finally, functional evenness reflects the uniformity of the distribution of abundances across the different functional types or positions in trait space. High functional evenness implies that functional roles are equally represented in terms of abundance.

A long-held view in the field posits that it is inherently difficult, if not impossible, to capture these three distinct facets within a single, unified index \citep{villeger2008, Mouchet2010}. This difficulty arises because each facet emphasizes different aspects of trait distribution. This perspective has historically led to the development of numerous, often disparate, metrics \citep{villeger2008, Mouchet2010, Schmera2023}, with even very recent proposals continuing this trend by introducing separate indices tailored to each aspect (e.g., NJ Trees Richness, NJ Trees Divergence, and NJ Trees Evenness \citep{Cardoso2024}). Furthermore, many existing metrics often overlook or struggle to adequately incorporate intraspecific trait variation (ITV), which is increasingly recognized as essential for a more nuanced and accurate quantification of FD \citep{palacio2024integrating}. These challenges underscore the urgent need for a new generation of FD metrics that can provide a more accurate, holistic, and theoretically sound assessment of ecological systems.

This paper proposes a solution to the challenges raised above. We demonstrate that the probability-weighted Vendi Score (pVS), first introduced by~\citet{friedman2023vendi}, can serve as a flexible and accurate FD index. We adapt pVS to functional ecology by defining it as the exponential of the Rényi entropy of the eigenvalues of the abundance-weighted trait similarity matrix. This formulation allows pVS to be applicable regardless of the unit of analysis. In particular, pVS can be defined at the individual level, allowing it to naturally incorporate intraspecific trait variation (ITV). This direct accounting for individual-level differences provides a more ecologically realistic and fine-grained assessment of FD than metrics relying solely on species-average traits, which often overlook crucial within-species variation. pVS can also be defined at the species level when individual-level data aren't available.

We mathematically prove that pVS rigorously satisfies several essential theoretical criteria for robust FD metrics. Notably, its adherence to properties such as set monotonicity (reflecting richness), distance monotonicity (reflecting divergence), and continuity in abundance (reflecting evenness) underscores its ability to holistically capture the multifaceted nature of FD. Furthermore, we demonstrate that pVS consistently exhibits the expected behavior across simulated ecosystem scenarios where ground-truth is known and where many existing FD metrics fail. 

\section{Theory}
Consider an ecological community (C), which is composed of a collection of $N$ observed biological individuals $(\bx_1, \dots, \bx_N)$. Each of these individuals is characterized by a unique set of functional traits, and we denote by $\bt_i$ the traits of $\bx_i$. These traits, such as body mass, leaf area, or metabolic rate, are measurable characteristics that define an organism's ecological role and influence its performance within the ecosystem. In our framework, each individual $\bx_i$ is also associated with an abundance $a_i$, which quantifies its presence or contribution within the community. This abundance can take various forms: it might be a biomass measure (e.g., the total mass of a fish), a cover measure (e.g., the percentage of ground covered by a plant), or even a measure reflecting energy use or metabolic rate. Denote by $p_i$ the relative abundance of $\bx_i$, calculated as the proportion of an individual's abundance relative to the total abundance of all individuals in the community, ensuring that all relative abundances sum to one. 

Once the traits and relative abundances of the members of $C$ are defined, we construct a similarity matrix $\bK \in \mathbb{R}^{N\times N}$ based on the pairwise trait distances between members. This matrix is generated using a positive semi-definite kernel similarity function $k(\cdot, \cdot)$ that is such that $k(\bx_i, \bx_j) = k(\bx_j, \bx_i)$ and $k(\bx_i, \bx_i) = 1$. These properties of the kernel are reflected in the similarity matrix $\bK$ which is defined such that $\bK_{ij} = k(\bx_i, \bx_j)$ and $\bK_{ii} = 1$. Note, the similarity function can be chosen depending on the application. For example, for quantitative traits, the Radial Basis Function (RBF) kernel with Euclidean distance is a common choice. Cosine similarity is another popular choice. For communities with a mix of quantitative and categorical traits, the Gower distance, or its generalizations \citep{Pavoine2009, dorazio2021gower_distances}, can be used to build the similarity matrix. Denote by ${\bK}_p$ the relative abundance-weighted similarity matrix,
\begin{equation}
    \bK_p = \text{diag}(\sqrt{\bp}) \, \bK \, \text{diag}(\sqrt{\bp})
\end{equation}
where $\text{diag}(\sqrt{\bp})$ is the diagonal matrix whose diagonal entries correspond to the square root of the relative abundances $\sqrt{\bp} = (\sqrt{p_1}, \dots, \sqrt{p_N})$.~This matrix ${\bK}_p$ effectively combines trait similarities with relative abundances. Let $\lambda_1, \dots, \lambda_N$ denote the eigenvalues of ${\bK}_p$. They reflect both functional similarity and abundance.

Applying the probability-weighted Vendi Score (pVS)~\citep{friedman2023vendi} to functional ecology, we define the functional diversity of $C$ as the exponential of the Renyi entropy of order $q$ of the eigenvalues,
\begin{align}\label{def:pVS}
    \text{pVS}(C, k, \bp, q) &= \exp\left(\frac{1}{1-q} \log \sum_{i=1}^{N} \lambda_i^q\right)
\end{align}
where we use the convention $0^0 = 0$ and $0\log 0 = 0$. This definition of pVS can also be applied when the unit of analysis is a species, by simply defining the abundances and the traits at the species level.

The order $q$ in Eq. \ref{def:pVS} determines the sensitivity of pVS to trait variations and abundance distribution within the community. For $q = 0$, pVS quantifies the effective number of functionally distinct groups within the community, essentially simply counting how many truly unique functional roles are present. When $0<q<1$, pVS is sensitive to rare functional types or individuals with unique trait combinations, even if they have low abundances, thereby giving more weight to the less dominant functions in the community. Conversely, when $q > 1$, pVS increasingly emphasizes dominant functional types or individuals with high abundance and common trait combinations, with pVS accounting for only the most prevalent functional role in the community as $q \rightarrow \infty$. Finally, when $q = 1$ (the limit as $q\rightarrow 1$), pVS corresponds to the Shannon entropy of the eigenvalues of ${\bK}_p$, providing a balanced measure that considers both the overall spread of functional types and the evenness of their contributions to the community, influenced by both their trait similarities and relative abundances. For the primary analyses in this paper, we primarily utilize $q=1$, for which pVS simplifies to:
\begin{align}\label{def:pVS_Shannon}
    \text{pVS}(C, k, \bp, 1) := \text{pVS}(C, k, \bp) &=  \exp \left(-\sum_{i=1}^N \lambda_i \log \lambda_i \right).
\end{align}
However, richer insights into a community's functional structure can be gained by considering the full spectrum of $q$ values. Indeed, two communities can achieve identical pVS values at $q=1$ yet achieve different pVS at other values of $q$ (see Fig. \ref{fig:q_behavior}). We therefore also demonstrate the flexibility of pVS by reporting \emph{functional diversity profiles}, which are curves representing pVS as a function of $q$. These profiles reveal how different facets of functional diversity (richness, divergence, or evenness) are emphasized at varying sensitivities to trait variations and abundance distributions. Functional divergence is captured by pVS across all $q$ values since pVS uses a trait similarity kernel. The $q$ parameter then modulates how this divergence is weighted alongside functional richness and evenness. For instance, at $q=0$, the profiles primarily reflect functional richness, highlighting the number of distinct functional types. As $q$ increases towards $q=1$, the profiles increasingly incorporate functional evenness, reflecting the regularity of abundance distribution across functional types. At $q>1$, the profiles become more sensitive to the evenness and divergence of the most dominant functional types. 

To better understand pVS's behavior for the specific value of $q = 1$ when all three facets are accounted for, we consider its response under extreme conditions of trait similarity and abundance distribution.

\parhead{Identical functional traits.}~If all members in the community possess identical functional traits, then the similarity matrix $\bK$ will be a matrix of all ones. In this scenario, the probability-weighted kernel matrix $\bK_p$ is a rank-1 matrix, with its only non-zero eigenvalue being $1$. Consequently, pVS will be minimally $1$. This correctly reflects a community with no functional diversity, regardless of the number of individuals or their abundance distribution.

\parhead{Maximally distinct functional traits.}~If all members have maximally distinct functional traits (i.e., they are as dissimilar as possible), the off-diagonal elements of $\bK_p$ will be zero. In this scenario, the eigenvalues of $\bK_p$ will be equal to the relative abundances $\bp$. The pVS is then simply the Hill number of order q of the community~\citep{hill1973diversity}. If abundances are uniform, pVS will be $N$, reflecting the maximum possible functional diversity for $N$ distinct individuals. 

\parhead{Uniform abundances.}~When all members of the community have uniform relative abundances ($p_i = \frac{1}{N}$ for all i), pVS simplifies to the Vendi Score~\citep{friedman2023vendi, pasarkar2023cousins}. Its value is then primarily driven by the functional dissimilarities among individuals, reflecting both functional richness and divergence, while inherently demonstrating high functional evenness due to the uniform abundances.

\parhead{Dominant species}.~If one community member is highly dominant (e.g., $p_i \approx 1$ and other $p_j \approx 0$ for $j \ne i$) then the $\bK_p$ matrix will be dominated by the contribution of that single species. Consequently, most eigenvalues will be very small, and pVS will approach 1, correctly indicating low functional diversity due to the overwhelming influence of a single functional type, irrespective of the total number of species or their potential functional distinctiveness.

We next highlight how pVS satisfies several essential theoretical criteria discussed in the functional diversity literature~\citep{solow1994, mason2003, ricotta2005}. All proofs are detailed in the appendix. 

\begin{theorem}[Invariance to functional redundancy]
\text{}
\begin{enumerate}   
\item \textit{Twinning.} Consider two communities $\mathcal{C}_A = (\bx_1, \dots, \bx_N)$  with relative abundance $\bp_A$ and $\mathcal{C}_B = (\bx_1, \dots, \bx_N, \bx_{N+1})$ with relative abundance $\bp_B$, where $\exists \text{ }i$ such that $k(\bx_i, \bx_{N+1}) = 1$. Then $\text{pVS}(\mathcal{C}_A, k, \bp_A, q) \geq \text{pVS}(\mathcal{C}_B, k, \bp_B, q)$.
\item \textit{Unaffected by the number of individuals/species.} Consider a community $\mathcal{C}_A = (\bx_1, \dots, \bx_N)$  with abundances $\bs_A = (s_1, \dots, s_N)$. Consider another community \\$\mathcal{C}_B = (\bx_1, \dots, \bx_N, \bx_{N}, \bx_{N+1}, \dots, \bx_{K(N+1)})$ for some K and such that $\bt_{N+kN+i} = \bt_i$ and $s_{k(N+1)+i} = s_i$. Define $\bp_A = (p_1, \dots, p_N)$ where $p_i = \frac{s_i}{\sum_{i=1}^{N} s_i}$. Define $\bp_B = (\tilde{p}_1, \dots, \tilde{p}_{K(N+1)})$ where $\tilde{p}_j = \frac{s_j}{\sum_{j=1}^{K(N+1)} s_j}$. Then $\text{pVS}(\mathcal{C}_A, k, \bp_A, q) = \text{pVS}(\mathcal{C}_B, k, \bp_B, q)$.
\item \textit{Invariance to species split.} Consider a community $\mathcal{C}_A = (\bx_1, \dots, \bx_N)$ with relative abundance $\bp_A$. Consider another community $\mathcal{C}_B = (\bx_1, \dots, \bx_{N-1}, \tilde{\bx}_N, \tilde{\bx}_{N+1})$ such that $\bt_N = \tilde{\bt}_N = \tilde{\bt}_{N+1}$ and $\bp_B = (\bp_{A,1}, \dots, \bp_{A,N-1}, \tilde{\bp}_N, \tilde{\bp}_{N+1})$ with $\bp_{A,N} = \tilde{\bp}_N + \tilde{\bp}_{N+1}$. Then $\text{pVS}(\mathcal{C}_A, k, \bp_A, q) = \text{pVS}(\mathcal{C}_B, k, \bp_B, q)$.
\end{enumerate}
\end{theorem}\label{thm:thm1}
This theorem underscores pVS's robustness as an FD metric by ensuring its invariance to various forms of functional redundancy. It guarantees that the index will not be inflated by the presence of functionally identical individuals or species (twinning), by an arbitrary increase in the number of taxa that do not introduce novel functional roles, or by arbitrary taxonomic decisions (species split) that do not alter the underlying functional composition. pVS consistently focuses on the unique functional roles present in the community, rather than simply counting taxa.

\begin{theorem}[Response to Community Composition Changes]\label{thm:thm3}
\text{}
\begin{enumerate}   
\item \textit{Set monotonicity.} Consider two communities $\mathcal{C}_A = (\bx_1, \dots, \bx_N)$ and $\mathcal{C}_B = (\bx_1, \dots, \bx_N, \bx_{N+1})$ with relative abundances $\bp_A$ and $\bp_B$, respectively. Assume $k(\bx_{N+1}, \bx_i) > 0$ \text{ } $\forall i$. Then $\text{pVS}(\mathcal{C}_A, k, \bp_A, q) \leq \text{pVS}(\mathcal{C}_B, k, \bp_B, q)$.
\item \textit{ Distance monotonicity.} Consider two communities $\mathcal{C}_A = (\bx_1, \dots, \bx_N)$ and $\mathcal{C}_B = (\tilde{\bx}_1, \dots, \tilde{\bx}_N)$ with relative abundances $\bp_A$ and $\bp_B$, respectively, such that $\bp_A = \bp_B$. Assume that $\forall $ $i,j \in \left\{1, \dots, N\right\}$ we have $k(\bx_i, \bx_j) \geq k(\tilde{\bx}_i, \tilde{\bx}_j)$. Then $\text{pVS}(\mathcal{C}_A, k, \bp_A, q) \geq \text{pVS}(\mathcal{C}_B, k, \bp_B, q)$.
\item \textit{ Sensitivity to functionally-rare species.} Consider a community $\mathcal{C}_A = (\bx_1, \dots, \bx_N)$. Assume $\exists$ i such that $k(\bx_i, \bx_j) \geq k(\bx_j, \bx_k)$ and $k(\bx_i, \bx_k) \geq k(\bx_j, \bx_k)$ $\forall$ $j \ne i$ and $k \ne i$. Denote by $m_i$ the abundance of $\bx_i$. Consider the community\\ $\mathcal{C}_B = (\bx_1, \dots, \bx_N)$ such that the abundance of $\bx_i$ is $m_i' \leq m_i$. Denote the relative abundances by $\bp_A$, $\bp_B$, respectively. Then $\text{pVS}(\mathcal{C}_A, k, \bp_A, q) \geq \text{pVS}(\mathcal{C}_B, k, \bp_B, q)$.
\item \textit{ Continuity in abundance.}  Consider a community $\mathcal{C}_A = (\bx_1, \dots, \bx_N)$. Denote by $w_i$ the relative abundance of $\bx_i$ for some $i$. Consider the community $\mathcal{C}_B = \mathcal{C}_A \setminus \left\{\bx_i\right\}$. Then $\lim_{w_i\to 0} \left|\text{pVS}(\mathcal{C}_A, k, q) - \text{pVS}(\mathcal{C}_B, k, q)\right| = 0$. 
\end{enumerate}
\end{theorem}
This theorem highlights how pVS dynamically responds to changes in community composition, directly reflecting the facets of functional richness, divergence, and evenness. Set Monotonicity ensures that the addition of novel functional roles (increasing functional richness) is appropriately captured as an increase in diversity. Distance Monotonicity confirms that greater functional dissimilarity among individuals or species (indicating higher functional divergence) translates to higher pVS values. Lastly, the last two properties related to abundance stated in the theorem demonstrate pVS's ability to account for functional evenness by weighing the contributions of individuals (or species) by their abundance, ensuring that dominant functional types have a greater impact and that the loss of rare, unique functions is reflected in the overall diversity score. The continuity property ensures that these responses are smooth and ecologically realistic.

\begin{theorem}[Fundamental Mathematical Properties]\label{thm:thm4}
\text{}
\begin{enumerate}   
\item \textit{Boundedness and positivity.} Consider a community $\mathcal{C}_A = (\bx_1, \dots, \bx_N)$ with relative abundance $\bp_A$. Then $1 \leq \text{pVS}(\mathcal{C}_A, k, \bp_A, q) \leq N$.
\item \textit{Concavity.} Consider $M$ non-overlapping communities $\mathcal{C}_1, \dots, \mathcal{C}_M$ such that for any $\bx \in \mathcal{C}_m$ and $\by \in \mathcal{C}_n$ we have $k(\bx, \by) = 0$ for all $n \ne m \in \left\{1, \dots, M\right\}$. Denote by $S_m$ the size of $\mathcal{C}_m$ and define $w_m = \frac{S_m}{\sum_{m'=1}^{M}}$. Denote by $\bp$ the relative abundance of the union of the communities and by $\bp_m$, the relative abundance of $\mathcal{C}_m$ for all $m$. Then $\text{pVS}\left(\bigcup_{m=1}^{M}\mathcal{C}_m, k, \bp, q\right) \geq \sum_{m=1}^{M} w_m\cdot \text{pVS}(\mathcal{C}_m, k, \bp_m, q)$.
\end{enumerate}
\end{theorem}
This theorem establishes some fundamental mathematical properties of pVS. Its positivity and boundedness between 1 and N highlight its interpretability, allowing pVS to be directly interpreted as an "effective number" of functional types. This facilitates straightforward comparisons across diverse communities, regardless of their raw richness. The concavity property is crucial for theoretical robustness. It implies that diversifying abundances (making them more even) will increase functional diversity in a predictable, diminishing-returns manner. 

Collectively, the properties stated in the theorems ensure that pVS is a well-behaved, interpretable, and theoretically sound metric that can reliably quantify all three facets of functional diversity: richness (through its effective number interpretation and response to distinct additions), divergence (through its sensitivity to dissimilarity), and evenness (through its response to abundance distributions and concavity).

\begin{figure}[!ht]
    \centering
    \includegraphics[width=\linewidth]{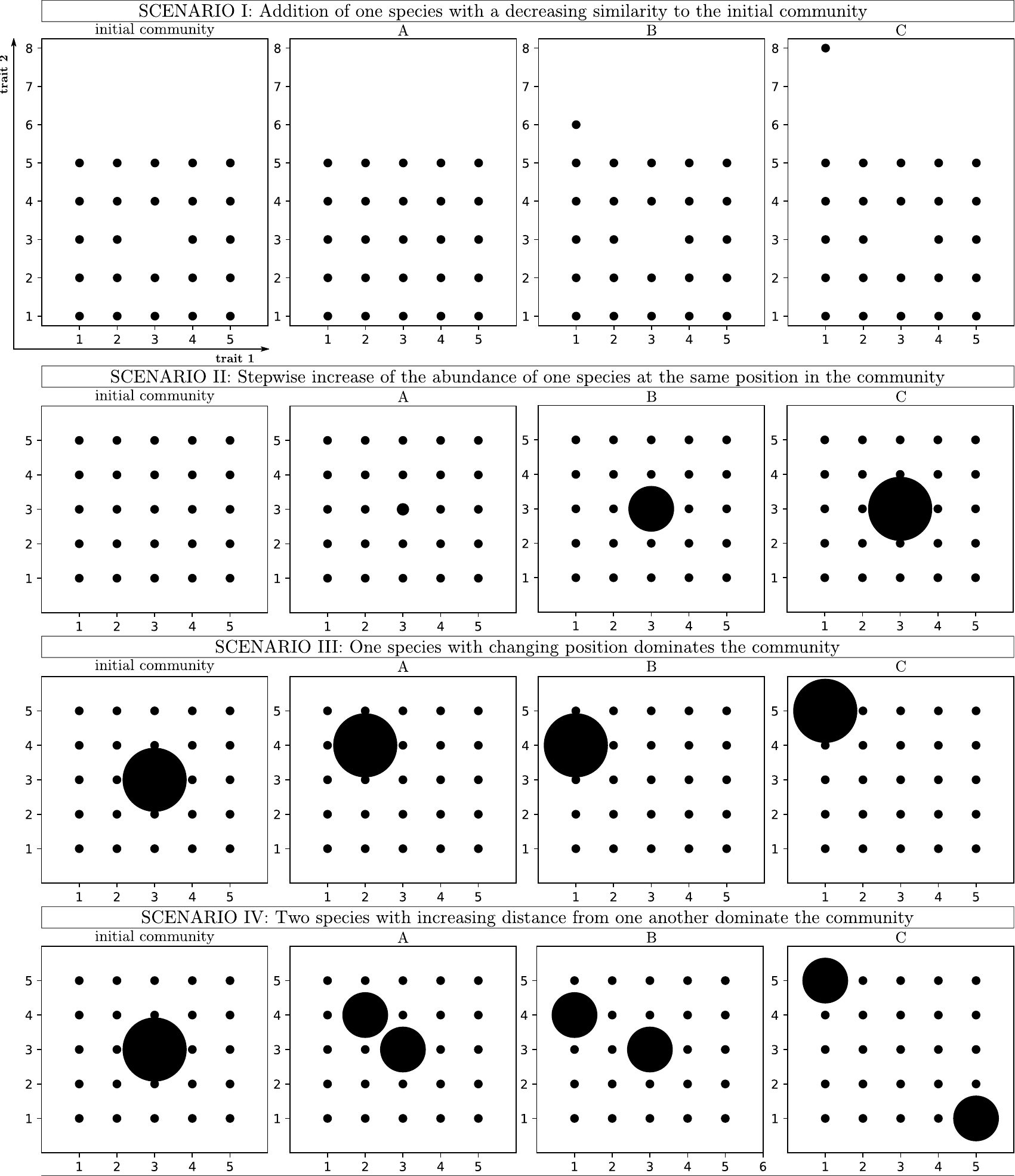}
    \caption{Simulated ecosystem scenarios with only two functional traits. A dot represents one species, and its size is correlated with its abundance.}
    \label{fig:experiment_dots}
\end{figure}

\section{Results}
To rigorously evaluate pVS and other FD indices, we utilized
the simulated scenarios detailed in \cite{Schleuter2010}. We illustrate these scenarios in Fig. \ref{fig:experiment_dots}. Only two functional traits are considered here, meaning that the functional trait space is 2-dimensional. Each species is represented by a black dot, and the size of the dot is correlated with the value of its abundance: a big dot means the abundance of the species is high, and a small dot means the abundance is small. The values of the traits are integers from 1 to 8. 

Scenario I corresponds to the case where one species with different trait values, and thus a different position in the functional trait space, is added to the community. Different values of similarity to the initial community are used while the abundances are kept equal for all species. Adding such a species to the community should increase functional diversity. We expect a higher increase when the similarity to the community is smaller, i.e., when the trait distance is higher. 

Scenario II consists in increasing the abundance of only one species in the community. We start with a balanced community (all abundances are equal), and we give more and more importance to one species. We expect a decrease in functional diversity as one species dominates in abundance. Indeed, an assemblage of 10 species dominated by one is less diverse than another with 10 equally represented species. 

Scenario III corresponds to the case where one species dominates the community in terms of abundance, but we now vary its position in the trait space. The similarity of the dominant species compared to the community varies. We expect an increase in functional diversity when the similarity decreases. 

In Scenario IV, two species dominate the community and have the same abundance. We increase the distance between these two species. The less similar they are, the more functionally diverse the community is supposed to be. 

\begin{table}[t]
\resizebox{\textwidth}{!}{%
\begin{tabular}{@{}lc}
\midrule
\textbf{Functional diversity index}                                               & \textbf{Validity Score} \\ \midrule
\texttt{Functional Attribute Diversity FaD}, [\cite{walker1999plant}]                                         & 1/4                  \\ \midrule
\texttt{modified Functional Attribute Diversity mFaD}, [\cite{schmera2009measuring}]                            & 1/4                  \\ \midrule
\texttt{Functional Diversity FD, [\cite{Petchey2002}]}                             & 1/4                  \\ \midrule
\texttt{Functional Richness, [\cite{villeger2008}]}                                & 1/4                  \\ \midrule
\texttt{Average contribution}, [\cite{ricotta2005note}] & 1/4                  \\ \midrule
\texttt{Rao's quadratic entropy, [\cite{bottadukat2005}]}                          & 4/4                  \\ \midrule
\texttt{Functional Divergence FDiv, [\cite{villeger2008}] }                        & 3/4                  \\ \midrule
\texttt{Functional Dispersion FDis, [\cite{laliberte2010}] }                       & 4/4                  \\ \midrule
\texttt{Index of Mason FDVar, [\cite{mason2003}] }                                 & 1/4                  \\ \midrule
\texttt{FRO index to multiple traits FEve, [\cite{villeger2008}]}                  & 1/4          \\ \midrule  
\texttt{Trait Onion Peeling TOP, [\cite{Fontana2015}]}                 & 1/4 \\ \midrule  
\texttt{Functional diversity index with NJ trees, Richness, [\cite{Cardoso2024}]}                 & 1/4 \\ \midrule 
\texttt{Functional diversity index with NJ trees, Dispersion, [\cite{Cardoso2024}]}                 & 3/4 \\ \midrule 
\texttt{Functional diversity index with NJ trees, Evenness, [\cite{Cardoso2024}]}                 & 3/4\\ \midrule
\texttt{pVS, [\cite{friedman2023vendi}]}                  & \textbf{4/4}\\  
\bottomrule
\end{tabular}%
}
\caption{Performance of different FD indices on the four simulated ecosystem scenarios. The validity score is computed by recording the number of scenarios in which a given index succeeds and dividing that number by $4$, the total number of scenarios. Most metrics fail in at least 1 scenario. The pVS, Rao's quadratic entropy, and FDis are the only metrics that succeed in all scenarios.}
\label{tab:scores}
\end{table}

Trait distance is computed using the $L_1$ distance. pVS uses a Radial Basis Function kernel to define similarity from these distances. The similarity kernel is
\begin{align}
    k(x,y)&=\exp\left(\frac{-d(x,y)^2}{2\sigma^2}\right)
\end{align}
where $d(x, y)$ is the distance between $x$ and $y$ and $\sigma^2$ is the width of $k$. 

\begin{figure*}[!hbpt]
    \centering
    \begin{subfigure}[b]{0.9\textwidth}
        \centering
        \includegraphics[width=\linewidth]{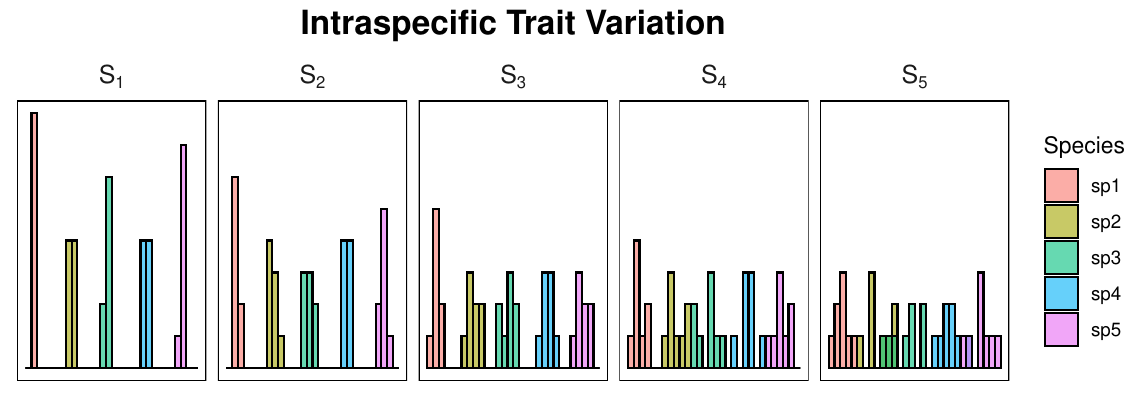}
    \end{subfigure}
    \begin{subfigure}[b]{0.28\textwidth}
        \centering
        \includegraphics[width=\linewidth]{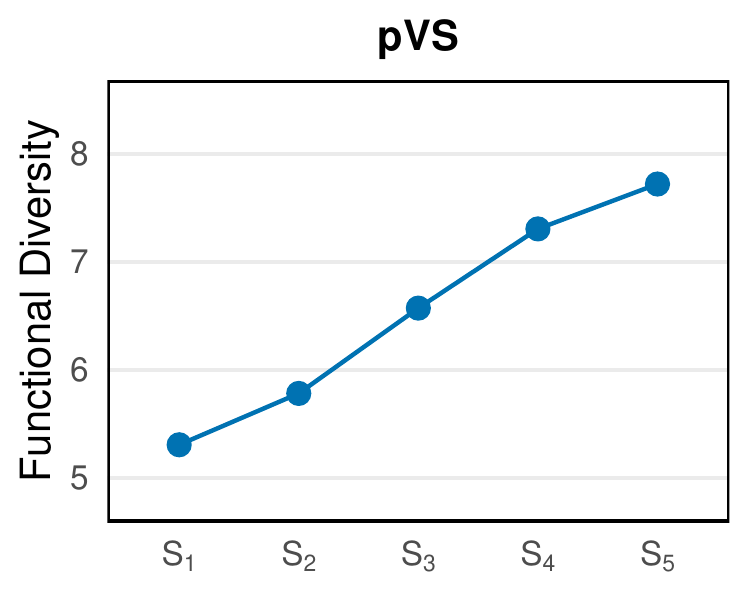}
    \end{subfigure}
    \begin{subfigure}[b]{0.28\textwidth}
        \centering
        \includegraphics[width=\linewidth]{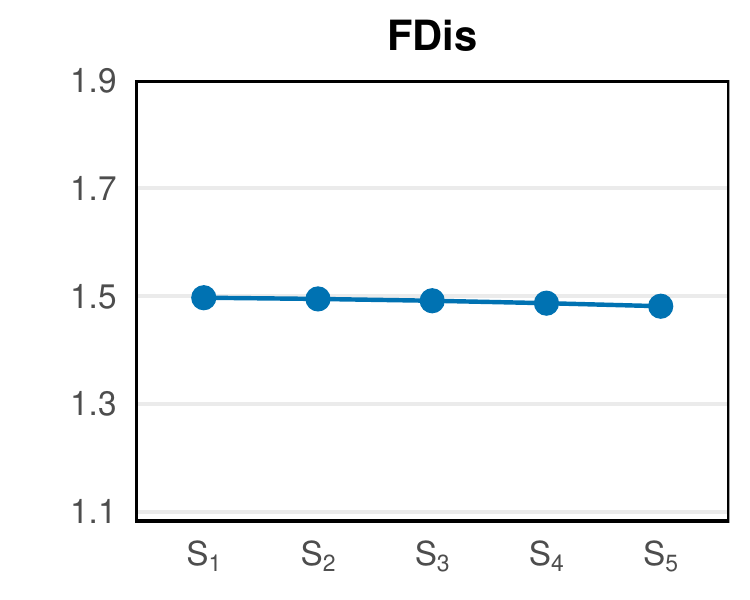}
    \end{subfigure}
    \begin{subfigure}[b]{0.28\textwidth}
        \centering
        \includegraphics[width=\linewidth]{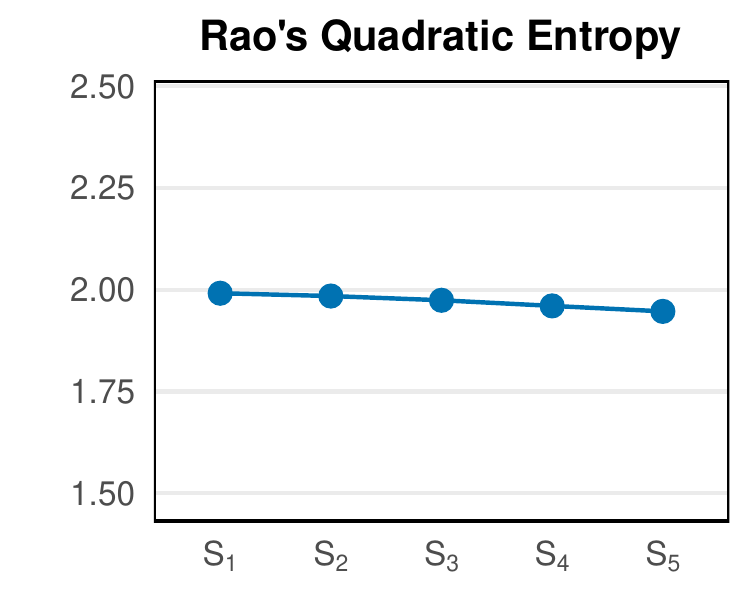}
    \end{subfigure}
    \begin{subfigure}[b]{0.28\textwidth}
        \centering
        \includegraphics[width=\linewidth]{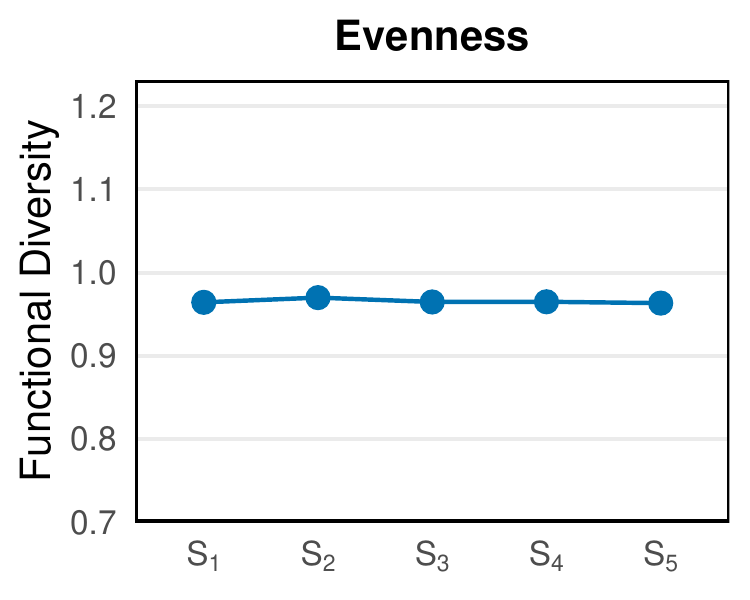}
    \end{subfigure}
    \begin{subfigure}[b]{0.28\textwidth}
        \centering
        \includegraphics[width=\linewidth]{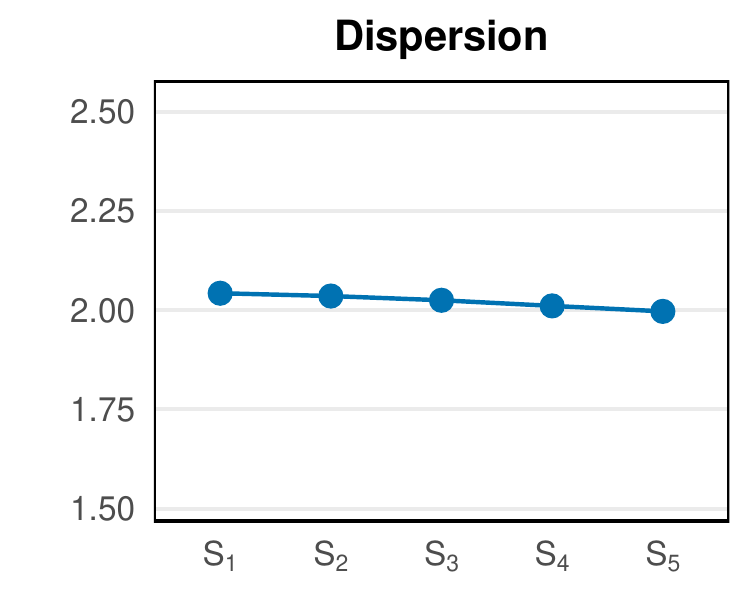}
    \end{subfigure}
    \begin{subfigure}[b]{0.28\textwidth}
        \centering
        \includegraphics[width=\linewidth]{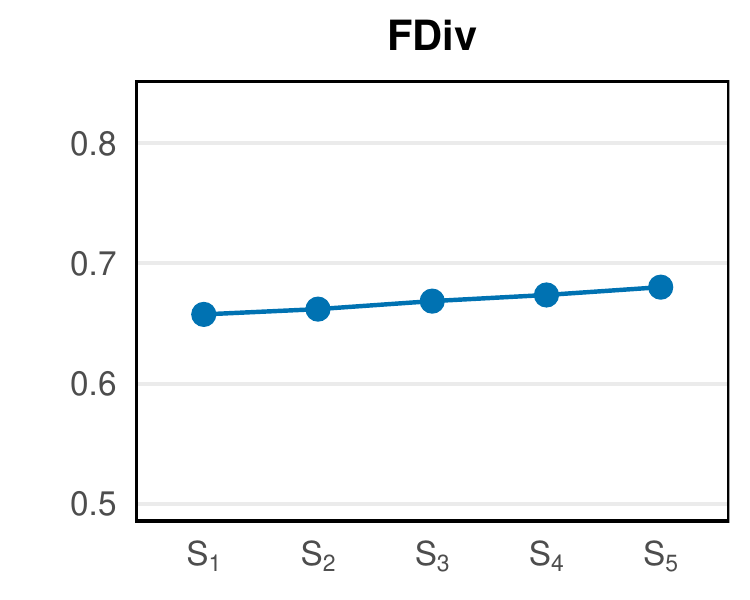}
    \end{subfigure}
    \begin{subfigure}[b]{0.28\textwidth}
        \centering
        \includegraphics[width=\linewidth]{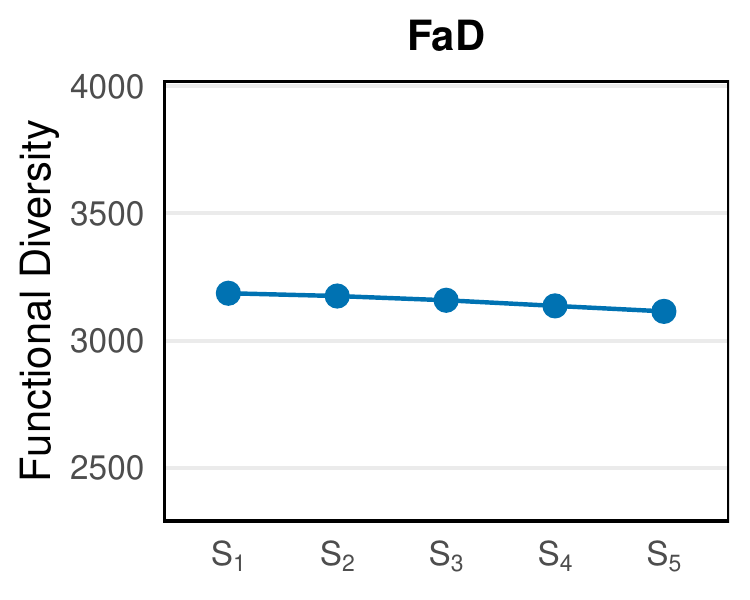}
    \end{subfigure}
    \begin{subfigure}[b]{0.28\textwidth}
        \centering
        \includegraphics[width=\linewidth]{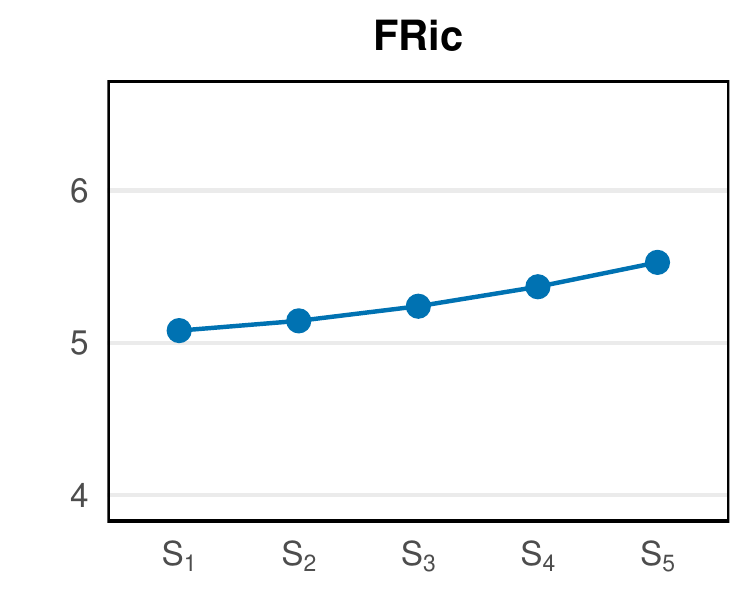}
    \end{subfigure}
    \begin{subfigure}[b]{0.28\textwidth}
        \centering
        \includegraphics[width=\linewidth]{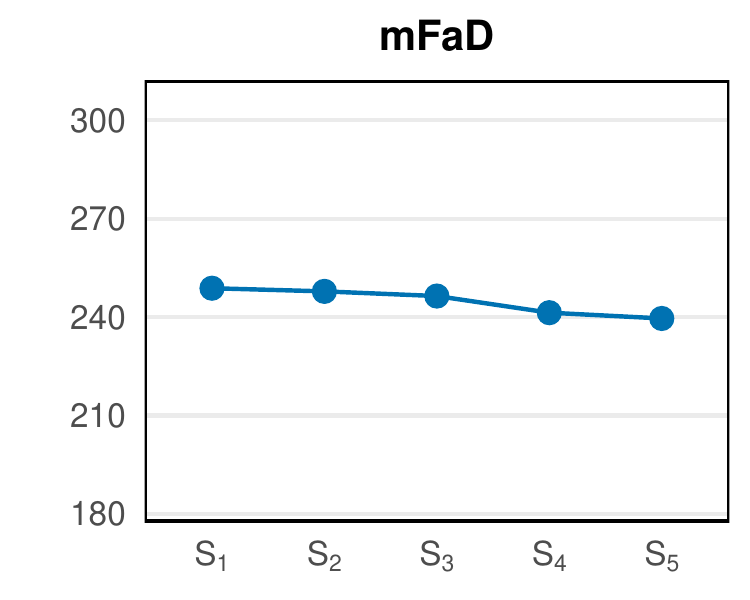}
    \end{subfigure}
    \begin{subfigure}[b]{0.28\textwidth}
        \centering
        \includegraphics[width=\linewidth]{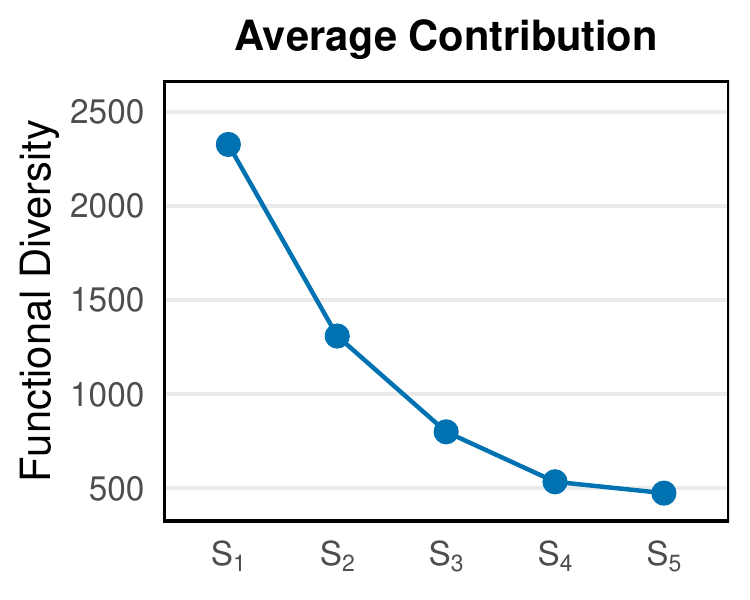}
    \end{subfigure}
    \begin{subfigure}[b]{0.28\textwidth}
        \centering
        \includegraphics[width=\linewidth]{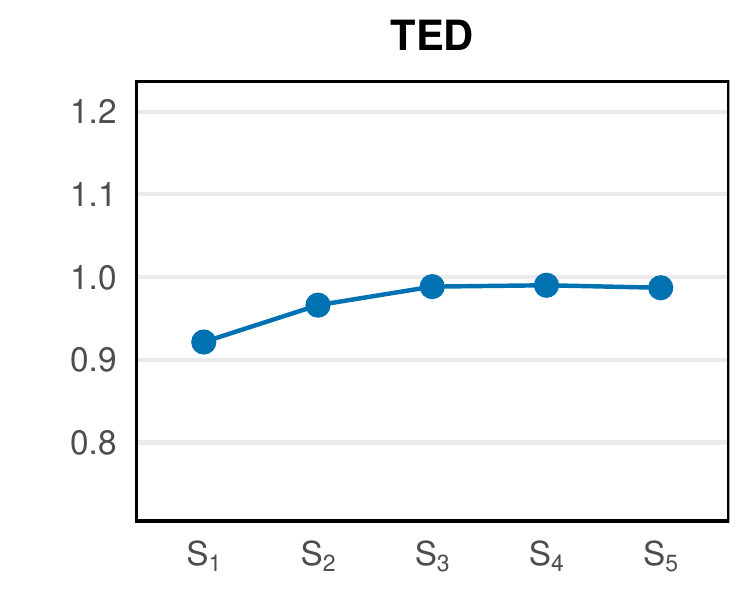}
    \end{subfigure}
    \begin{subfigure}[b]{0.28\textwidth}
        \centering
        \includegraphics[width=\linewidth]{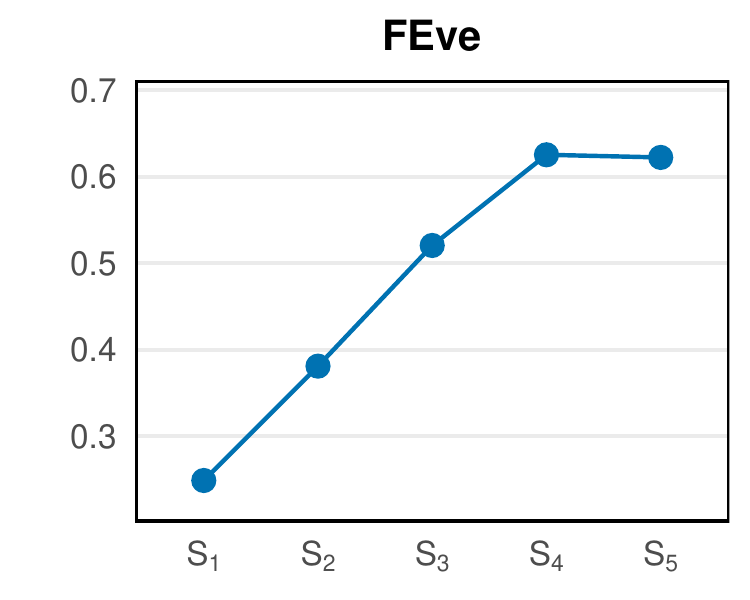}
    \end{subfigure}
    \begin{subfigure}[b]{0.28\textwidth}
        \centering
        \includegraphics[width=\linewidth]{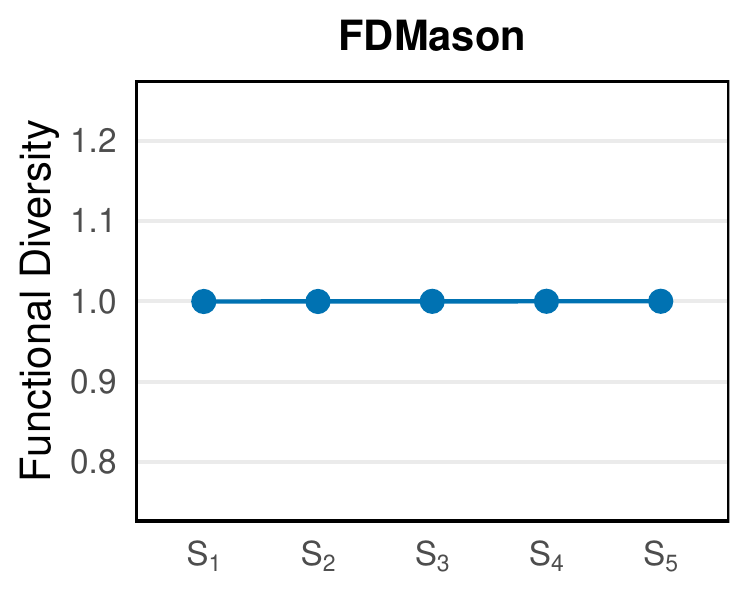}
    \end{subfigure}
    \begin{subfigure}[b]{0.28\textwidth}
        \centering
        \includegraphics[width=\linewidth]{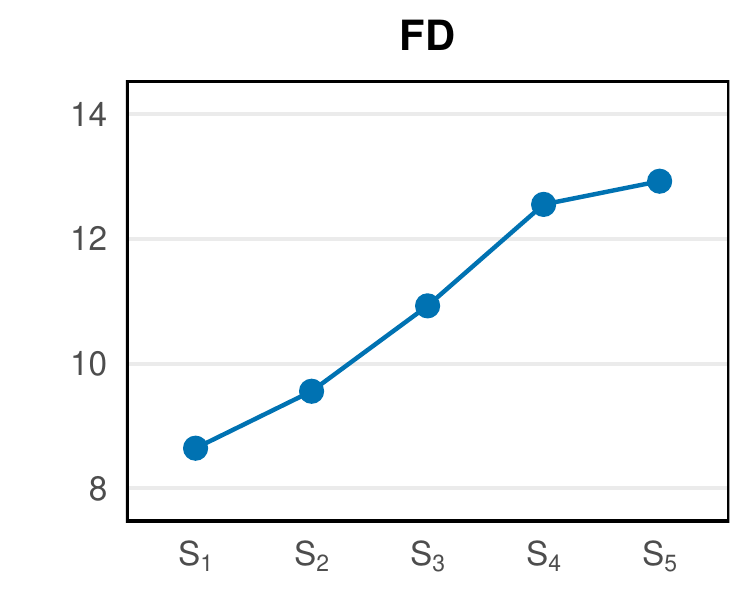}
    \end{subfigure}
    \begin{subfigure}[b]{0.28\textwidth}
        \centering
        \includegraphics[width=\linewidth]{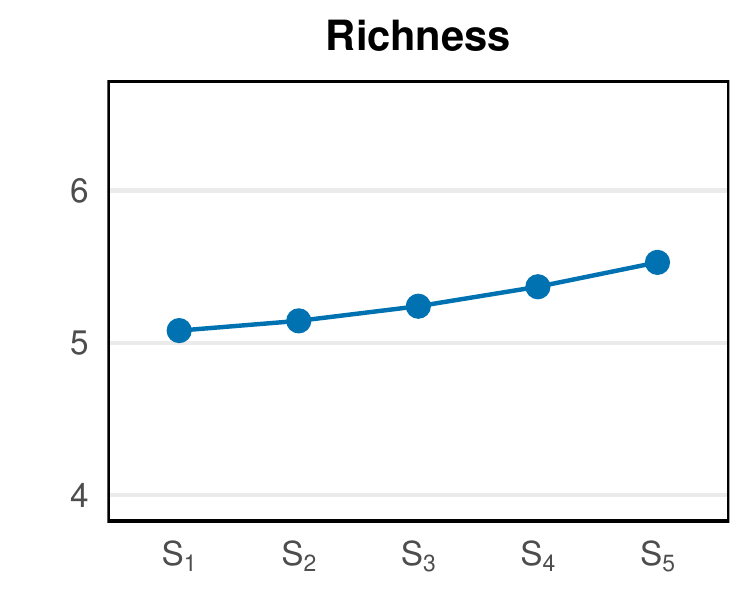}
    \end{subfigure}
    \caption{\textbf{Top:} Histograms of trait values for synthetic communities with increasing levels of variation from left to right. \textbf{Bottom:} Behavior of 15 different FD indices, computed at the individual level, as intraspecific trait variance increases. Only pVS, FDiv, FRic, FDMason, FD, and Richness increase monotonically across the $5$ communities. FDMason and FDiv both increase by less than 0.1\% and 4\% between community $S_1$ and community $S_5$ despite the drastic changes to the community's traits. Rao's quadratic entropy and FDis fail to capture ITV. pVS is the only metric that is both accurately sensitive to ITV and passes all scenarios from Table \ref{tab:scores}.} 
    \label{fig:ITV}
\end{figure*}

To quantify performance, we compute a \emph{validity score} for each index. We define the validity score as the number of scenarios in which the index exhibited the expected ecological response to community changes divided by the total number of scenarios. As shown in Table \ref{tab:scores}, 9 out of the 15 FD indices only succeed in 1 scenario and fail on all the others. On the other hand, 3 out of the 15 metrics fail on 1 scenario, with Rao's quadratic entropy \cite{ricotta2005}, Functional Dispersion (FDis)~\citep{laliberte2010}, and pVS being the only indices that succeed in all scenarios. 

We additionally show that pVS can account for intraspecific trait variation (ITV), unlike the metrics that pass the majority of scenarios in Table \ref{tab:scores}, i.e. Rao's quadratic entropy and Functional Dispersion (FDis). In Figure \ref{fig:ITV}, we simulate communities of 5 species with $1$ trait, and increasing the levels of trait variance for each species. Intuitively, we expect functional diversity to increase when we introduce more intraspecific variance, yet most metrics are either constant or decreasing. Among those that are sensitive to ITV, pVS is the only one to pass all scenarios in Table \ref{tab:scores}. (Note: we omit the TOP metric here because it requires at least two-dimensional trait data.)

\begin{figure*}[!hbpt]
    \centering
    \includegraphics[width=\linewidth]{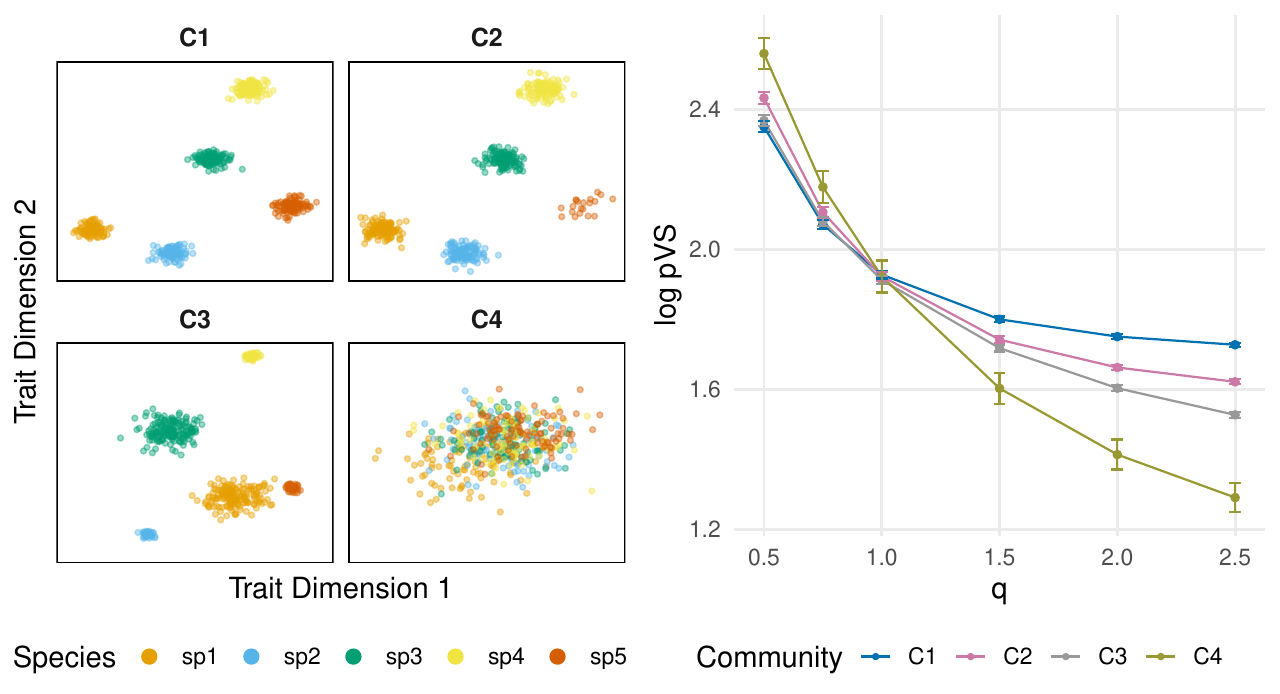}
    \caption{The order q of pVS prioritizes distinct community characteristics. \textbf{Left:} 4 communities consisting of 5 species with 2 traits. C1 contains $5$ species of equal size. $C2$ has $1$ rare species and slightly higher variance within the other $4$ species. $C3$ has two dominant species. $C4$ has 5 similar species that all have high variance. \textbf{Right:} the \emph{\textbf{FD profile}} of the four different communities as described by the curve of log-pVS as a function of q across $10$ simulations. All communities have almost the same pVS at $q=1$ (difference within 1 standard deviation), but have pVS that differ significantly at other values of $q$. Community C4 has the highest pVS at q=0.5, but the lowest at larger values of $q$. Community C1 has the lowest pVS at q=0.5, but the highest at larger values of $q$.} 
    \label{fig:q_behavior}
\end{figure*}
Finally, we empirically show how the order $q$ used in the pVS will prioritize different aspects of functional diversity. In Figure \ref{fig:q_behavior}, we highlight $4$ examples of these differences. When $q<1$, the pVS places greater emphasis on the presence of rare species and the overall spread of trait values, leading to higher pVS for communities with broader overall trait distributions, like communities $C2$ and $C4$. As $q$ increases, the pVS becomes increasingly sensitive to the size of dominant species, which explains why communities with more even abundances across species have higher pVS, like community C1. In contrast, communities with many highly similar trait values, like community C4, tend to drop in rank. Community $C3$ has two distinct dominant species, giving it a higher pVS than $C4$ for large q. Tuning $q$ allows researchers to focus on either rarity and dispersion (low $q$) or dominance and functional redundancy (high $q$), making pVS a flexible tool for studying various ecological communities.

\section{Conclusion}
The robust and comprehensive quantification of functional diversity (FD) is a long-standing challenge in ecology. Existing metrics struggle to simultaneously capture richness, divergence, and evenness, and often neglect intraspecific trait variation (ITV). In this paper, we demonstrate that the probability-weighted Vendi Score (pVS) can be adapted to functional ecology to provide a robust and widely-applicable FD index. Defined as the exponential of the Rényi entropy of the eigenvalues of the abundance-weighted trait similarity matrix, pVS can be applied at the individual level, which naturally accounts for ITV and provides a more fine-grained FD assessment. It is also readily applicable at the species level by aggregating individual data. We rigorously prove that pVS satisfies several essential theoretical criteria for robust FD metrics and consistently exhibits the expected ground-truth behavior across simulated ecosystem scenarios where many existing functional diversity metrics fail. By unifying the multifaceted nature of FD and incorporating ITV within a single, theoretically sound framework, pVS offers a powerful, interpretable, and comprehensive tool for ecological research. 

\section*{Code and Data Availability}
The Python code to compute the probability-weighted Vendi Score (pVS) on your data is freely available in the Vertaix GitHub at \url{https://github.com/vertaix/Vendi-Score}. The pVS is also available as a Python package at \url{https://pypi.org/project/vendi-score/}. The pVS is also readily available in R in the same GitHub at \url{https://github.com/vertaix/Vendi-Score-R}. 

All the results in this paper can be reproduced using the R code in the repo \url{https://github.com/vertaix/pVS-Functional-Diversity}. That same repo contains the code to generate the data for the ecosystem scenarios and reproduce Table 1. 

\bibliographystyle{apa}
\bibliography{References}

\begin{thebibliography}{}

\bibitem[\protect\astroncite{Botta-Dukát}{2005}]{bottadukat2005}
Botta-Dukát, Z. (2005).
\newblock Rao's quadratic entropy as a measure of functional diversity based on multiple traits.
\newblock {\em Journal of Vegetation Science}, 16(5):533--540.

\bibitem[\protect\astroncite{Cardoso et~al.}{2024}]{Cardoso2024}
Cardoso, P., Guillerme, T., Mammola, S., Matthews, T.~J., Rigal, F., Graco‐Roza, C., Stahls, G., and Carlos~Carvalho, J. (2024).
\newblock Calculating functional diversity metrics using neighbor‐joining trees.
\newblock {\em Ecography}, 2024(7).

\bibitem[\protect\astroncite{D'Orazio}{2021}]{dorazio2021gower_distances}
D'Orazio, M. (2021).
\newblock Distances with mixed type variables some modified gower's coefficients.

\bibitem[\protect\astroncite{Fontana et~al.}{2015}]{Fontana2015}
Fontana, S., Petchey, O.~L., and Pomati, F. (2015).
\newblock Individual‐level trait diversity concepts and indices to comprehensively describe community change in multidimensional trait space.
\newblock {\em Functional Ecology}, 30(5):808–818.

\bibitem[\protect\astroncite{Friedman and Dieng}{2023}]{friedman2023vendi}
Friedman, D. and Dieng, A.~B. (2023).
\newblock The vendi score: A diversity evaluation metric for machine learning.
\newblock {\em arXiv preprint arXiv:2210.02410}.

\bibitem[\protect\astroncite{Hill}{1973}]{hill1973diversity}
Hill, M.~O. (1973).
\newblock Diversity and evenness: a unifying notation and its consequences.
\newblock {\em Ecology}, 54(2):427--432.

\bibitem[\protect\astroncite{Laliberté and Legendre}{2010}]{laliberte2010}
Laliberté, E. and Legendre, P. (2010).
\newblock A distance‐based framework for measuring functional diversity from multiple traits.
\newblock {\em Ecology}, 91(1):299–305.

\bibitem[\protect\astroncite{Mason et~al.}{2003}]{mason2003}
Mason, N.~W., MacGillivray, K., Steel, J.~B., and Wilson, J.~B. (2003).
\newblock An index of functional diversity.
\newblock {\em Journal of Vegetation Science}, 14(4):571–578.

\bibitem[\protect\astroncite{Mouchet et~al.}{2010}]{Mouchet2010}
Mouchet, M.~A., Villéger, S., Mason, N. W.~H., and Mouillot, D. (2010).
\newblock Functional diversity measures: an overview of their redundancy and their ability to discriminate community assembly rules.
\newblock {\em Functional Ecology}, 24(4):867–876.

\bibitem[\protect\astroncite{Palacio et~al.}{2024}]{palacio2024integrating}
Palacio, F.~X., Ottaviani, G., Mammola, S., Graco-Roza, C., de~Bello, F., and Carmona, C. (2024).
\newblock Integrating intraspecific trait variability in functional diversity: An overview of methods and a guide for ecologists.

\bibitem[\protect\astroncite{Pasarkar and Dieng}{2023}]{pasarkar2023cousins}
Pasarkar, A.~P. and Dieng, A.~B. (2023).
\newblock Cousins of the vendi score: A family of similarity-based diversity metrics for science and machine learning.
\newblock {\em arXiv preprint arXiv:2310.12952}.

\bibitem[\protect\astroncite{Pavoine et~al.}{2009}]{Pavoine2009}
Pavoine, S., Vallet, J., Dufour, A., Gachet, S., and Daniel, H. (2009).
\newblock On the challenge of treating various types of variables: application for improving the measurement of functional diversity.
\newblock {\em Oikos}, 118(3):391–402.

\bibitem[\protect\astroncite{Petchey and Gaston}{2002}]{Petchey2002}
Petchey, O.~L. and Gaston, K.~J. (2002).
\newblock Functional diversity (fd), species richness and community composition.
\newblock {\em Ecology Letters}, 5(3):402–411.

\bibitem[\protect\astroncite{Ricotta}{2005a}]{ricotta2005}
Ricotta, C. (2005a).
\newblock A note on functional diversity measures.
\newblock {\em Basic and Applied Ecology}, 6(5):479–486.

\bibitem[\protect\astroncite{Ricotta}{2005b}]{ricotta2005note}
Ricotta, C. (2005b).
\newblock A note on functional diversity measures.
\newblock {\em Basic and applied Ecology}, 6(5):479--486.

\bibitem[\protect\astroncite{Schleuter et~al.}{2010}]{Schleuter2010}
Schleuter, D., Daufresne, M., Massol, F., and Argillier, C. (2010).
\newblock A user’s guide to functional diversity indices.
\newblock {\em Ecological Monographs}, 80(3):469–484.

\bibitem[\protect\astroncite{Schmera et~al.}{2009}]{schmera2009measuring}
Schmera, D., Podani, J., and Er{\H{o}}s, T. (2009).
\newblock Measuring the contribution of community members to functional diversity.
\newblock {\em Oikos}, 118(7):961--971.

\bibitem[\protect\astroncite{Schmera et~al.}{2023}]{Schmera2023}
Schmera, D., Ricotta, C., and Podani, J. (2023).
\newblock Components of functional diversity revisited: A new classification and its theoretical and practical implications.
\newblock {\em Ecology and Evolution}, 13(10).

\bibitem[\protect\astroncite{Solow and Polasky}{1994}]{solow1994}
Solow, A.~R. and Polasky, S. (1994).
\newblock Measuring biological diversity.
\newblock {\em Environmental and Ecological Statistics}, 1(2):95–103.

\bibitem[\protect\astroncite{Villéger et~al.}{2008}]{villeger2008}
Villéger, S., Mason, N. W.~H., and Mouillot, D. (2008).
\newblock New multidimensional functional diversity indices for a multifaceted framework in functional ecology.
\newblock {\em Ecology}, 89(8):2290–2301.

\bibitem[\protect\astroncite{Walker et~al.}{1999}]{walker1999plant}
Walker, B., Kinzig, A., and Langridge, J. (1999).
\newblock Plant attribute diversity, resilience, and ecosystem function: the nature and significance of dominant and minor species.
\newblock {\em Ecosystems}, 2(2):95--113.

\end{thebibliography}

\end{document}